ARTICLE

# Early turbulence and pulsatile flows enhance diodicity of Tesla's macrofluidic valve

Quynh M. Nguyen [1,2], Joanna Abouezzi[1] & Leif Ristroph [1 ✉]

Microfluidics has enabled a revolution in the manipulation of small volumes of fluids. Controlling flows at larger scales and faster rates, or *macrofluidics*, has broad applications but involves the unique complexities of inertial flow physics. We show how such effects are exploited in a device proposed by Nikola Tesla that acts as a diode or valve whose asymmetric internal geometry leads to direction-dependent fluidic resistance. Systematic tests for steady forcing conditions reveal that diodicity turns on abruptly at Reynolds number $\mathrm{Re} \approx 200$ and is accompanied by nonlinear pressure-flux scaling and flow instabilities, suggesting a laminar-to-turbulent transition that is triggered at unusually low $\mathrm{Re}$. To assess performance for unsteady forcing, we devise a circuit that functions as an AC-to-DC converter, rectifier, or pump in which diodes transform imposed oscillations into directed flow. Our results confirm Tesla's conjecture that diodic performance is boosted for pulsatile flows. The connections between diodicity, early turbulence and pulsatility uncovered here can inform applications in fluidic mixing and pumping.

[1] Applied Math Lab, Courant Institute of Mathematical Sciences, New York University, New York, NY, USA. [2] Department of Physics, New York University, New York, NY, USA. ✉email: ristroph@cims.nyu.edu

From circulatory and respiratory systems to chemical and plumbing networks, controlling flows is important in many natural settings and engineering applications[1–5]. Perhaps the simplest means for directing flows is through the geometry of vessels, pipes, channels, and networks of such conduits. How geometry maps to flow patterns and distribution, however, is a challenging problem that depends on flow regime, as characterized by dimensionless quantities involving length- and time-scales and fluid material properties[6]. The field of microfluidics focuses on low Reynolds numbers in which small volumes are conveyed at slow speeds, and recent progress stems from advances in microscale manufacturing and lab-on-a-chip applications[7]. The flow physics at these scales is dominated by pressures overcoming viscous impedance, and the linearity of the governing Stokes equation enables theoretical and computational approaches that greatly aid in the design of microfluidic devices[8]. At larger scales and faster rates, the applications are as numerous and important[2,3] but the flow physics quite different. The underlying Navier–Stokes equation is nonlinear, theoretical results are fewer, simulations are challenging, and the mapping between geometry and desired fluidic objectives all the more complex[6]. The phenomenology of high-Reynolds-number or inertia-dominated flows is well documented: Flows are slowed in boundary layers near surfaces and tend to separate in a manner sensitive to geometry to yield vortices, wakes, jets, and turbulence[6,9]. Such complexities are exemplified by the breakdown of reversibility: Running a given system in reverse, say by inverting pressures, does not in general cause the fluid to move in reverse but instead triggers altogether different flow patterns[6].

Here we explore how the physics of inertial flows is exploited in an intriguing device proposed by the inventor Nikola Tesla[10]. Tesla intended this 'valvular conduit'—an image of which is reproduced from his 1920 patent in Fig. 1a—to allow fluid to pass easily in one direction while presenting substantially higher resistance to flows in the reverse direction. Such a fluidic diode or valve can be used as a fundamental component for directing flows. While it is unclear if Tesla ever fabricated and tested a prototype, its unique geometry of linked and looped lanes has attracted many studies into its operation and potential applications[11–41]. Previous fluid mechanical tests have focused on steady conditions of constant flow rate, confirming the resistance asymmetry or diodicity at high Reynolds numbers (Re) for geometries modified from Tesla's original design[11–15]. While it is clear that valving action is absent at low Re ≪ 1 for which flows are reversible[39] but permissible for sufficiently high Re > 1[13–21], there is lacking a characterization spanning flow regimes that would reveal how diodicity varies over a wide range of Re. Consequently, it is unclear if and how diodicity relates to internal flow phenomena such as the laminar-to-turbulent transition[6,9]. Further, Tesla's patent emphasizes unsteady or pulsatile flow conditions[10], which arise in pumping or rectification applications[11–13,15,17,24,33] but for which the potential for enhanced diodic performance remains to be determined.

In this work, we aim to experimentally test the valvular conduit across a wide range of steady and unsteady conditions. On a conduit whose planform shape is faithful to Tesla's original design, we first carry out systematic characterizations of flow resistance or friction for fixed pressure differences. Following measurement and analysis procedures established for flows through long, slender pipes and ducts, our experiments reveal abrupt changes in fluidic properties that are reminiscent of the laminar-to-turbulent transition, albeit triggered at anomalously low Re. We then seek to evaluate Tesla's conjecture[10] that diodic performance is strongest when flow "is supplied in pulses and, more especially, when the same is extremely sudden and of high frequency." We propose and implement a fluidic circuit or network subject to unsteady forcing via imposed oscillatory flows, which the diodic conduits transform or rectify into one-way flows, the whole system serving as an AC-to-DC converter or pump. A quasi-steady model links these two studies by using the steady-forcing characteristics to predict the response for unsteady forcing. Our results indicate that pulsatile flows significantly enhance pumping performance, thus bearing out Tesla's conjecture and suggesting optimal operating conditions for rectifying conduits more generally.

## Results

**Experimental tests under steady forcing.** We first characterize experimentally the fluidic resistance or flow-induced pressure losses for Tesla's diode under conditions of fixed pressure differences, this quantity varied systematically to explore flow rates in both directions. We realize a conduit whose planform or overhead-view geometry is faithful to Tesla's original design[10], and we pursue Reynolds numbers ranging over orders of magnitude, the latter being important for our later comparison of steady versus unsteady (oscillatory) forcing. We digitize the planform of Fig. 1a and manufacture a macroscale version of the conduit via laser-cutting and bonding clear acrylic plastic, a 3D rendering of which is shown in Fig. 1b. We select a scale that, together with the use of water and water-glycerol mixtures as the working fluids, allow for characterization of the channel across low to high Re. The overall length is $L = 30$ cm, average wetted width $w = 0.9$ cm and depth $d = 1.9$ cm.

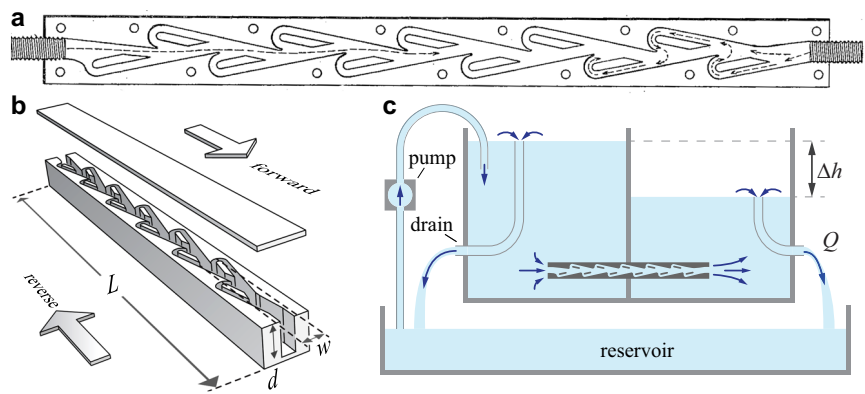

**Fig. 1 Experimental tests of Tesla's conduit under steady pressures. a** Schematic modified from Tesla's patent[10] showing a planform view of the 'valvular conduit'. **b** Rendering of the conduit used in experiments. Upper and lower lids sandwich the internal geometry, which is digitized from Tesla's design, laser-cut, and bonded. Relevant dimensions include total length $L$, average wetted width $w$, and depth $d$. **c** Schematic of the pressure chamber. Overflow mechanisms ensure fixed water levels that drive flow through the conduit, whose actual orientation is upright as shown in **b**. The height differential $\Delta h$ is varied and volumetric flow rate $Q$ measured for both forward and reverse directions.

To impose and controllably vary the pressure difference across the channel, we design and construct the apparatus whose sectional view is shown in Fig. 1c. Two chambers of a tank are connected only via the conduit, and the liquid level in each can be set and stably maintained via overspill mechanisms. The level difference $\Delta h$ is set by two adjustable internal drains whose heights can be independently varied. The hydrostatic pressure difference across the channel is $\Delta p = \rho g \Delta h$, where $\rho$ is the density of the fluid and $g = 980$ cm/s$^2$ is gravitational acceleration[6]. Liquid flows from the high side to the low side through the channel at a volumetric flow rate $Q$ and out to the reservoir at the same rate. A pump takes fluid from the reservoir, slightly overflowing the high side and thus maintaining its level. The system is closed and runs indefinitely. In this way, a pressure difference may be imposed and recorded by measuring the column heights with rulers, and the volumetric flow rate $Q$ is measured by intercepting the lower drain with a beaker of known volume and reading the fill-up time with a stopwatch. The flow direction is changed by simply changing which chamber has higher level.

The measured flow rate $Q$ versus $\Delta h$ for pure water is shown in Fig. 2a. As expected, increasing the height difference yields higher flow rates for both the forward and reverse directions, as defined in Fig. 1b. The flow rate $Q$ increases monotonically but nonlinearly with $\Delta h$. Importantly, for the same $\Delta h$, $Q$ is greater for the forward direction than reverse across all values of $\Delta h$. This anisotropy is more clearly seen in Fig. 2b, where the resistance $R = \Delta p/Q$ is plotted versus $Q$ for the forward and reverse directions. Across all $Q$, the resistance in reverse is greater, and this disparity increases with $Q$.

The errors for the data of Fig. 2a and b, as determined by multiple measurements at each condition, are smaller than the symbols and have been suppressed in these plots. Errors in $\Delta h$ are about a millimeter due to the height of the meniscus that obscures the reading of water level. Errors in $Q$ are set by the reaction time in triggering a stopwatch after collecting a specified volume. Large volumes and long collection times ( > 60 s) ensure relative errors under 1%.

**Pressure drop, friction and diodicity across flow regimes.** Experiments carried out with pure water yield high Reynolds numbers Re = $\rho U D/\mu \sim 10^3$, where $\mu$ is the fluid viscosity, $U = Q/wd$ is the section-averaged flow speed through the channel, and $D = 4V_w/S_w = 0.8$ cm is its hydraulic diameter calculated from the total wetted volume $V_w$ and the wetted surface area $S_w$. (The latter generalizes the conventional form $D = 4A_w/P_w$ for a conduit whose cross-section shape is uniform and of wetted area $A_w$ and perimeter length $P_w$[42].) To explore lower Reynolds numbers, we conduct further experiments with water-glycerol solutions of varying viscosity, and all results are combined in Fig. 2c. To account for the different fluid properties, we plot non-dimensional pressure or Hagen number $Hg = (\Delta p/L)(D^3\rho/\mu^2)$ versus non-dimensional flow rate or Reynolds number[6,43]. As expected, the disparity between the forward and reverse directions is significant only at sufficiently high Re. Also shown for comparison are reference lines indicating linear and quadratic scalings of pressure with flow rate. For low Re, it is seen that $Hg \sim Re$, which is expected for well-developed and laminar flow[6]. For higher Re, the data follow a stronger dependence of approximately $Hg \sim Re^2$, which is characteristic of turbulent flow[6]. Interestingly, the disparity in resistance occurs together with the nonlinearity of the Hg-Re scaling.

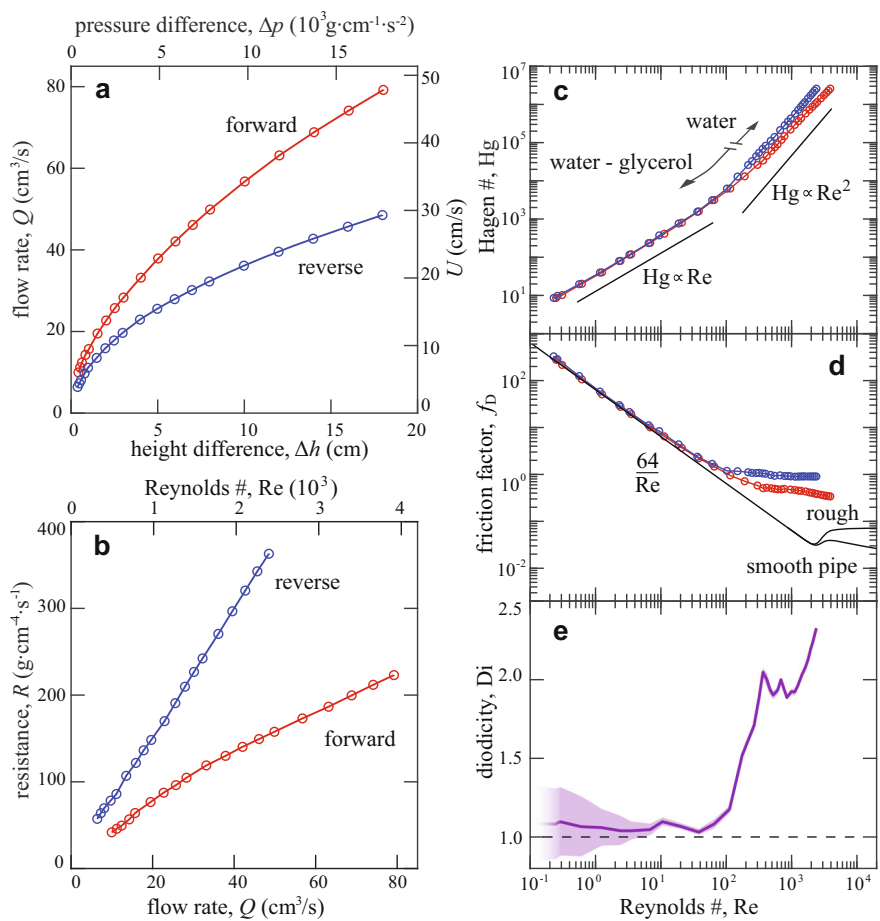

**Fig. 2 Experimental characterization of Tesla's conduit under steady pressures. a** Flow rate $Q$ across varying pressure heads $\Delta h$ and pressure differences $\Delta p = \rho g \Delta h$ for the case of pure water as the working fluid. The forward (red) and reverse (blue) directions exhibit different $Q$ for the same $\Delta p$. Here and elsewhere, error bars are suppressed when smaller than the symbol size (see text). **b** Hydrodynamic resistance $R = \Delta p/Q$ versus $Q$ for the forward and reverse directions. **c** Dimensionless forms of pressure difference (Hagen number Hg) versus flow rate (Reynolds number Re). The plot combines data on pure water and water-glycerol solutions to cover a wide range of Re. **d** Friction factors $f_D = (\Delta p/L)/(\rho U^2/2D)$, a dimensionless form of pressure drop appropriate for turbulent flow. Also shown are curves representing previous measurements for smooth and rough pipes. **e** Diodicity Di or ratio of reverse to forward resistances versus Re. The band represents propagated standard errors determined from repeated measurements.

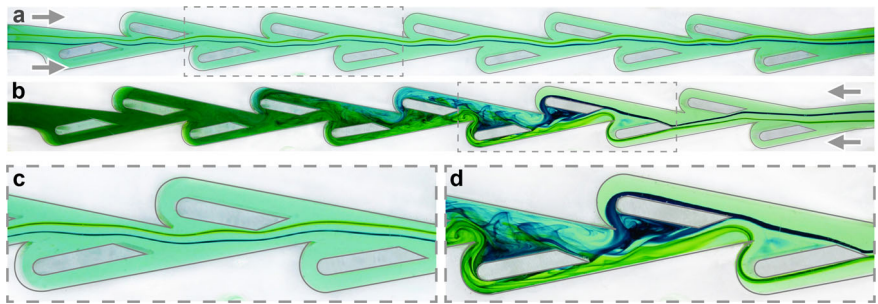

**Fig. 3 Streakline flow visualization at Re = 200 using dye injected upstream. a**, **c** Forward direction. Two adjacent filaments remain in the central corridor of the conduit with only small lateral deflections. **b**, **d** Reverse direction. The filaments ricochet off the periodic structures, deflecting increasingly sharply before being rerouted around the `islands' and mixing.

A conventional nondimensionalization of resistance used in studies of pipe and channel flow is the Darcy friction factor $f_D = (\Delta p/L)/(\rho U^2/2D)$, which normalizes pressure drop based on inertial scales[42,44]. In Fig. 2d we plot our measurements of $f_D(Re)$ for forward and reverse flow through Tesla's conduit. For comparison, we include on this so-called Moody diagram previous results on circular pipes[45]. The two curves shown correspond to smooth pipes and one of high roughness in which wall variations measure 10% of the mean diameter. The form $f_D = 64/Re$ corresponds to the Hagen–Poiseuille law and applies well to both smooth and rough-walled pipes in the laminar flow regime of $Re < 2000$[46]. Following a transitional region, well-developed turbulence tends to be triggered at higher $Re > 4000$, for which $f_D$ is more constant with Re and increases with roughness. By comparison, Tesla's conduit transitions away from the laminar-flow scaling at significantly lower $Re \approx 100$. Further, the order-one friction values at higher Re are substantially higher than those for turbulent flow through smooth and rough pipes, which reflects the high impedance presented by the complex geometry.

In interpreting these results, it is important to note that the association between the scaling of $Hg(Re)$ or $f_D(Re)$ with flow state (laminar or turbulent) holds only for sufficiently long, slender conduits. For short pipes and channels, entrance effects due to the developing flow can lead to turbulent-like ($Hg \sim Re^2$ and $f_D \sim Re^0$) rather than laminar-like ($Hg \sim Re$ and $f_D \sim Re^{-1}$) scaling even for laminar flow[6,42]. To ensure entrance effects are negligible for turbulent flows, the length-to-diameter ratio is typically recommended to exceed about 40[47], which is nearly satisfied by the value $L/D = 38$ for Tesla's conduit. For laminar flow, the aspect ratio should exceed the (dimensionless) entrance length of approximately $Re/30$[6], which is satisfied for $Re \lesssim 1000$ for the valvular conduit. These estimates suggest that the results reported here are representative of sufficiently slender geometries for which the pressure drop scaling can be linked to flow state, and the inclusion of our measurements on the standard Moody diagram of Fig. 2d is warranted.

The performance of the channel as an asymmetric resistor can be quantified by its *diodicity* or ratio of reverse to forward resistance values[48]. Equivalently, we define this ratio using dimensionless forms of pressure drops at the same Re: $Di(Re) = Hg_R(Re)/Hg_F(Re) = f_{D,R}(Re)/f_{D,F}(Re)$, where the subscripts indicate the reverse (R) and forward (F) directions. In Fig. 2e the curve indicates how Di varies with Re, with the band representing propagated errors based on repeated measurements. For low Re, Di is close to unity and remains so up until $Re \approx 100$. Over a narrow transitional range $Re = 100-300$, the diodic function of the channel abruptly 'turns on' or is activated, and for $Re = 300-1500$ we find $Di \approx 2$. Future work should investigate the behavior for $Re > 2000$.

Interestingly, the turn-on of diodicity apparent in Fig. 2e comes along with the nonlinear scaling of pressure drop with flow rate (Fig. 2c) and the departure from the laminar-flow friction law (Fig. 2d). These results suggest that diodic function is closely linked to a transition to turbulent flow that occurs significantly earlier (at lower Re) than that observed for smooth and rough pipes.

**Flow visualization and early turbulence**. Towards understanding the mechanisms behind these observations, we next visualize the internal flows in the conduit. We focus first on a transitional value of $Re = 200$, for which we inject neutrally buoyant dye upstream and record photographs and video using a camera positioned to view the planform. The conduit is clear and backlit, and the resulting images reveal flow streaklines[6]. Two adjacent streaklines near the middle of the channel are color-coded using blue and green dyes. Figure 3a shows the case of flow in the forward direction. The streaklines remain in the central corridor along the entire length of the channel and are deflected only slightly as they pass the periodic structures. Details of the gently meandering path can be seen in the zoomed-in image of Fig. 3c. In contrast, the reverse direction involves amplified lateral deflections of the streams that eventually result in strong mixing, as shown in Fig. 3b. The incoming filaments ricochet off the internal structures, with the redirections being only slight in passing the first 'islands' or baffles but quickly growing downstream after repeated interactions. The flows are eventually rerouted into the recesses, and the fluid is well mixed by the end of the channel. Some of the steps that destabilize the initially-laminar flow can be seen in the zoomed-in view of Fig. 3d.

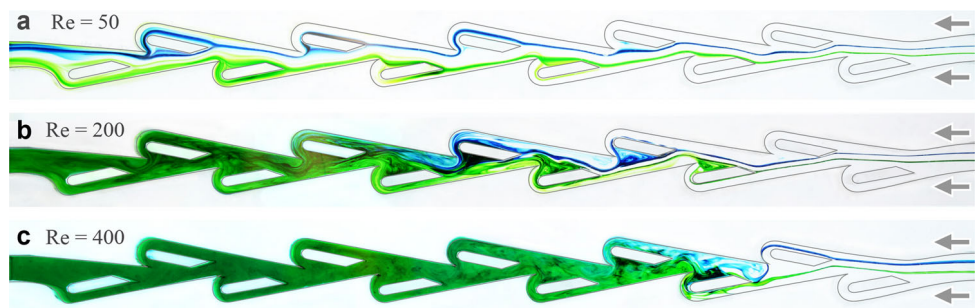

**Fig. 4 Transition in reverse flow state with increasing Reynolds number. a** Streakline visualization at Re = 50. Blue and green dyed filaments disperse but do not intermix, and the flow is steady throughout the conduit. **a** At a transitional value of Re = 200, the filaments are laminar and steady for the first few units, unsteady and intermix in the middle, and are nearly completely mixed by the end of the conduit. **c** At Re = 400, unsteady and well-mixed flows appear throughout most of the channel.

We next aim to link the transition in resistance and turn on of diodicity to changes in flow state for different Reynolds numbers. In Fig. 4 we compare the reverse flows visualized at Re = 50, 200 and 400, corresponding to conditions just before, during and just after the turn-on, respectively. For Re = 50, dye filaments remain on their respective sides of the conduit, dispersed by interactions with the islands but not intermixing. The flows are observed to be laminar and steady throughout the entire channel. For Re = 400, unsteadiness of the streaklines is apparent beyond the first units, after which the filaments rapidly combine over a few units to yield well-mixed flows over most of the length. Comparatively, the transitional state of Re = 200 displays a hybrid of these features: The filaments are steady and laminar over the first 3 or 4 units, become unsteady and cross sides, and then reach near complete mixing by the end.

These results confirm the high-Re irreversibility reported in previous studies, which have emphasized the circuitous route taken by reverse flows[21,22,25]. Our visualizations reveal the nature of the reverse flow instability and also the extent of mixing, which we associate with increased dissipation and resistance. Unsteady flows and increased resistance are hallmarks of turbulent flow, which is triggered for Re in the thousands for pipe and channel flow[6,49]. Our visualizations of flow destabilization in Tesla's conduit at considerably lower Re = 200 offer further evidence for an early transition to turbulence triggered by the complex geometry.

In interpreting this early turbulence phenomenon, a concern may be that the Reynolds number defined here based on the mean speed inadequately captures the local flow conditions at various positions in the conduit. However, close inspection of the reverse flows in the Supplementary Video indicates that speeds at different sites along the central and diverted lanes are comparable to one another, with differences measuring less than 50%. Hence, the onset of turbulence at unusually low Re ≈ 200 cannot be attributed to local surges in flow speed significant enough to reach the conventional transitional value of Re ≈ 2000 for pipe flow. An alternative interpretation for the early onset of turbulence is given in our concluding discussions.

**Unsteady forcing of a fluidic AC-to-DC converter.** Having characterized Tesla's conduit for steady pressure differences, we next consider unsteady forcing in which the internal flows are driven to oscillate. To assess Tesla's conjecture of enhanced performance for pulsatile flows[10], we draw on the analogy between electric and fluidic circuits and consider a full-wave rectifier that uses four diodes arranged in a bridge configuration in order to convert an imposed alternating current (AC) in one branch into directed current (DC) in a second branch[50]. The electric circuit is shown schematically in Fig. 5a. An AC current source is on the left, and the directionality of each ideal diode is indicated by the arrowhead. These elements are linked by conducting wires, and current directions are shown in red and blue for the two half-cycles of the AC source. When current is driven upwards through the source, only the two diodes under favorable bias conduct, and the current follows the red path. In the next half-cycle, the other pair of diodes conducts, and the current follows the blue path. Thus, while the input branch is purely AC or oscillatory, the output branch on the bottom exhibits a DC component or non-zero mean.

Figure 5b shows the schematic of the fluidic analog that we design, construct and test. Laser-cut and bonded Tesla conduits serve as diodes, a reciprocating piston replaces the AC current source, and these elements are connected in bridge configuration through piping. The circuit is filled with water, and the position of the piston is driven sinusoidally in time with amplitude $A$ and frequency $f$ controllably varied via a high-torque stepper motor (Longs Motor) and Arduino controller. Because the piston completely seals the surrounding cylinder, the flow in the AC branch is purely oscillatory. The diodic behavior of the conduits then manifest as one-way or directed current (DC) in the lower branch. To assess this, we use Particle Image Velocimetry (PIV) to measure the flow velocity field along a segment of the transparent DC branch pipe. The 5-cm-long interrogation region is encased in a rectangular water jacket to minimize optical distortion[51]. Following procedures from earlier studies[52–54], we seed the water with particles (hollow glass microspheres of approximate diameter 50 $\mu$m, 3M) whose near neutral buoyancy is ensured by selection from a fractionation column in water. A laser sheet (1.25W CW green, CNI) of thickness 0.5 mm is shone across the mid-plane along the PIV section, and the resulting particle motions are recorded via high-speed camera (12 MP, 150 fps, Teledyne Dalsa Falcon2). Post-processing via an established PIV algorithm[55], these data provide the flow velocity profile across the pipe, resolved in time within each oscillation cycle and over a total duration of at least 10 cycles.

Representative data are shown in Fig. 5c for one set of $A$ and $f$. The upper panel presents the flow speed averaged across the cross-section in the AC branch, where the sinusoidal oscillations $2\pi A f \sin(2\pi f t)$ conform to the piston motions. The lower panel represents the measured section-averaged flow speed in the DC branch, and the inset shows the flow velocity profile furnished by PIV at two points in the cycle. At each time $t$, the two halves of the profile (bisectioned by the tube's axis) are averaged and axisymmetry is assumed to arrive at the section-averaged speed. Strikingly, the flow has a dominant DC component $U_{DC}$, and the flow profiles remain unidirectional throughout the oscillation cycle. Thus the circuit achieves the goal of AC-to-DC conversion or pumping. The output flow also shows a weak AC component of amplitude $\Delta U$. These ripple oscillations occur at twice the driving frequency $f$, as both half-strokes of the AC input contribute to the DC output.

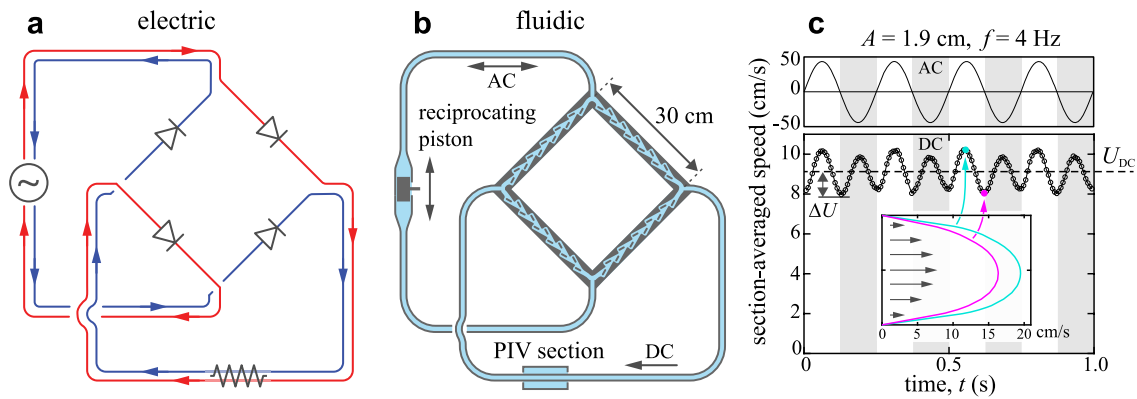

**Fig. 5 Electronic AC-to-DC converter and an analogous fluidic pump.** a Electric circuit with four ideal diodes that converts alternating current source (AC, left branch) to direct current (DC, lower branch). Red and blue lines highlight the path and direction of current at different phases in the AC cycle. b Sectional view of an analogous fluidic circuit with four Tesla diodes and a pulsatile flow source. The experimental device employs a reciprocating piston of amplitude $A$ and frequency $f$ as an AC source in one branch, and the DC flow is measured in a second branch. c Section-averaged flow imposed in the AC branch (top) and measured in the DC branch (bottom) for $A = 1.9$ cm and $f = 4$ Hz. The mean flow rate $U_{DC} > 0$ indicates successful AC-DC conversion or pumping. Inset: sample flow velocity profiles measured by PIV.

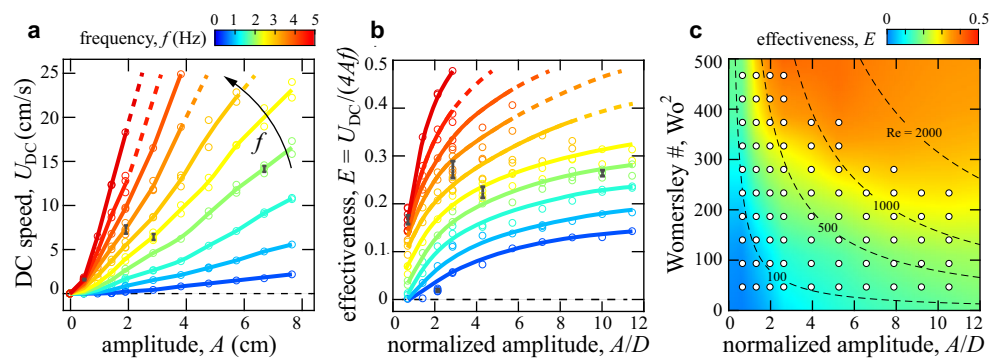

**Fig. 6 Performance of the fluidic AC-to-DC converter or pump.** a Average DC flow speed $U_{DC}$ versus driving amplitude $A$ and at different frequencies $f$. Representative error bars (black) show standard errors of the mean. b Effectiveness of rectification versus amplitude normalized by the hydraulic diameter $D$. c Experimentally measured effectiveness. The axes correspond to normalized amplitude and non-dimensional frequency or Womersley number $Wo^2 \propto f$. White markers indicate data points, and the colormap is interpolated and extrapolated elsewhere. The dashed hyperbolic curves are contours of the driving or oscillatory Reynolds number $Re \propto Wo^2 \cdot A/D$.

To assess the pumping performance of the circuit more generally, we systematically vary the AC input parameters $A$ and $f$ and measure the section-averaged DC flow speed $U_{DC}$, which is equivalent to volumetric flux (volume per unit cross-sectional area and time). Figure 6a shows how $U_{DC}$ varies with $A$ and $f$. Across all driving conditions, $U_{DC} > 0$ and the system achieves AC-to-DC conversion. As expected, the output $U_{DC}$ increases with the inputs $A$ and $f$. Less apparent in Fig. 6a is that the response is nonlinear. To clarify this, we define the *effectiveness* of the pump as $E = U_{DC}/4Af$. This normalization is chosen such that ideal or perfect diodes yield $E = 1$: A volume of fluid proportional to the piston displacement $2A$ is injected into the DC branch in each stroke of duration $1/2f$, with the two strokes in each cycle contributing equally. In Fig. 6b we plot effectiveness $E$ versus frequency $f$ and dimensionless amplitude $A/D$. The fact that $E < 1$ for all conditions reflects the non-ideal nature or 'leakiness' of the diode. Interestingly, it is seen that $E$ itself increases with both $A$ and $f$, quantifying the nonlinear response of $U_{DC}$. That is, doubling either $A$ or $f$ leads to disproportionately higher $U_{DC}$. For the conditions studied here, we achieve $E \approx 0.5$, and the trends suggest yet higher efficacy would be attained for stronger driving.

A fully dimensionless characterization is displayed in Fig. 6c, where $E$ is mapped across varying $A/D$ and the (squared) Womersley number $Wo^2 = \pi \rho f D^2/2\mu$, which assesses the unsteadiness of pulsatile flow by comparing frequency to the timescale for diffusion of momentum[56]. The high values of $Wo^2 = 50$ to 500 explored here suggest plug-like flow profiles in the AC sections. The variations in the map again highlight the nonlinearity of the pump, which is most effective in the red region of high $Wo^2 \propto f$. For reference, we include contours (dashed hyperbolic curves) of constant driving or oscillatory Reynolds number, defined as $Re = \rho A f D/\mu = 2/\pi \cdot Wo^2 \cdot A/D$. Significant pumping of $E > 0.1$ occurs for $Re$ in the hundreds, when diodicity is observed to turn on for steady flow (Fig. 2e).

**A quasi-steady model of the AC-to-DC converter.** The rectifying circuit provides a clean context for assessing Tesla's conjecture of enhanced performance of the diode for pulsatile flows. Our strategy involves formulating a model that predicts the pump rate of the system based on its steady-flow characteristics, and then comparing this prediction to the actual performance measured experimentally. The quasi-steady model views the AC–DC rectifier as a network of nonlinear resistors whose resistance values vary with flow rate and direction as measured and characterized in Fig. 2. The network can then be analyzed by standard methods for electronic circuits, i.e. by solving for unknown currents through all segments via equations for current/flow conservation at each node or junction and voltage/pressure drops around closed loops[50].

Complete model equations and calculations are available in the Methods section, and here we highlight the key assumptions and steps. We seek the instantaneous current or volumetric flux $Q(t)$ through each segment of the circuit. The resistance-current curves for each diode are given by fitting splines to the data of Fig. 2, where the sign of $Q$ in each diode dictates whether the forward or reverse resistance applies. The DC branch presents a constant resistance $R_{DC}$ estimated via the Hagen–Poiseuille law for pipe flow[6]. The AC source imposes $Q_{AC} = 2\pi f A w d \sin(2\pi f t)$ across the bridge. For any resistive element, pressure drops and currents are related via Ohm's law $\Delta p = QR$, with all quantities being time dependent. Kirchhoff's laws demand that pressure drops sum to zero around each closed loops and currents sum to zero at every node. Symmetry arguments reduce the unknowns to the DC branch current $Q_{DC}$ and two diode currents, for example those through the rightmost pair in Fig. 5b. One loop law and two node laws give three nonlinear algebraic equations for these unknown currents at each time $t$. Discretizing in time and applying MATLAB's *fsolve* function yields numerical solutions for the instantaneous currents. The effectiveness predicted by the model is then $E_M = <Q_{DC}(t)>/<|Q_{AC}(t)|>$, where the brackets indicate averages over a period.

**Comparing steady and unsteady performance**. The model furnishes predictions across varying inputs $A$ and $f$, and these results serve as a quasi-steady baseline against which the measured performance under unsteady conditions can be compared. In the colormap of Fig. 7a, we plot the so-called *boost* or relative enhancement of the experimentally measured effectiveness over the model prediction: $B = E/E_M$. The axes are again dimensionless forms of amplitude $A/D$ and frequency $Wo^2$. Warmer colors with $B > 1$ indicate conditions for which the actual circuit outperforms quasi-steady expectations. It can be seen that the device performs better than expected for all but the lowest values of $A$ and $f$, providing validation of Tesla's conjecture of enhanced diodic performance for pulsatile flows[10]. Further, unsteady effects seem to be optimally exploited for low amplitude and high frequency oscillations (red region), for which we observe boosts as high as $B \approx 2.5$ and thus more than a doubling in pump rate over the quasi-steady baseline. Extrapolation of these data suggests yet greater enhancement for higher frequencies.

Another point of comparison between the model and experiment involves the oscillations or ripples apparent in the DC branch signal, an example from experiments shown in the lower panel of Fig. 5c. We define the *pulsatility* $P = \Delta U/U_{DC}$ as the ratio of the ripple amplitude to the mean pump rate, which can be assessed from the experimental measurements and from the model output. In both cases, we fit the form $U_{DC} + \Delta U \sin(2\pi f t + \phi)$ to the section-averaged flow speed to extract the necessary quantities. Smooth output flows and thus low $P$ are generally preferable in pumping applications. In the quasi-steady model, we observe uniformly high $P_M \approx 1$ for all driving conditions (data not shown). This behavior is similar to an electronic diode-bridge rectifier, whose output current reaches a minimum of zero whenever the source current crosses zero, leading to oscillations as large as the mean. As shown in the map of Fig. 7b, the actual fluidic circuit proves to be much smoother with $P \approx 0.1$ across the conditions explored here. Thus the fluctuations are an order of magnitude less than that predicted by the quasi-steady model. Surprisingly, the DC output in experiments is less pulsatile for stronger AC driving, and high $A$ in particular yields low $P$. This effect and the general mitigation of pulsatility as compared to quasi-steady expectations may be due to flow inertia, which tends to filter out fluctuations but which is absent from the quasi-steady framework. This hypothesis might be explored in models or simulations that include inertia.

## Discussion

This work presents systematic experimental characterizations of Nikola Tesla's fluidic valve or diode across a wide range of both steady and unsteady flow conditions. The case of steady pressure/flow-rate considered in previous studies on Tesla-like channels provides a point of comparison to our results[13,16–18,21–24,31,34,35,13]. No earlier work reports on the abrupt rise in diodicity, which likely reflects the singular values or narrow ranges in Re explored. Further, previous experiments and simulations have reported weaker diodicity for corresponding conditions[11–13,15,18], e.g. Di = 1.0 – 1.3 at Re = 500 for which we measure Di = 2.0. This may be due to differences in geometry, both in that the diodes are variations on Tesla's original design and the measurement systems may not isolate the channel, thereby introducing additional pressure drops that dilute diodicity. The more encouraging results reported here are in better alignment with values Di = 2 – 4 reported[14,15,17,18,20] for various Tesla-inspired geometries and at select Re values between $10^3$ and $10^6$. This regime of Re > 2000 would benefit from a systematic characterization of diodicity as presented here for Re < 2000. Future research might also include shape optimization towards improving performance[12–15,17,18,18,19,24,34], though it seems that the diodicity value of 200 stated in Tesla's patent[10] seems well beyond reach, at least for the fluid dynamical regime studied here. Tesla may have envisioned air or other gases as the working fluid, in which case compressibility may affect performance but only at the extreme flow rates needed to achieve Mach numbers of order one.

Previous experimental and computational visualizations for Tesla-inspired channels have confirmed the reversible and irreversible flows expected at low[39] and high Re[21,22,25], respectively. Studies of the latter validate Tesla's picture of the reverse flows taking a circuitous path around the periodic internal structures[10]. Our work bridges flow regimes by focusing on the transitional Re = 200 and thereby revealing the destabilization mechanism for the reverse mode: Repeated interactions of the central flow with these structures lead to amplified lateral deflections that reroute streamlines and eventually induce complete mixing. These flow observations together with our resistance characterizations point to the conclusion that Tesla's conduit induces a turbulent-like flow state at unusually low Re. Namely, the signatures of pipe flow turbulence that first appear in smooth and rough-walled pipes at Re ≈ 2000 are triggered in Tesla's conduit at Re ≈ 200. Signs of such early turbulence, though with less pronounced changes in the transitional Re, have been observed for textured channels and corrugated pipes[46,57,58]. Our measurements establish a link between early turbulence and the turn-on of diodicity, a connection that should be tested in other asymmetric conduits[59–62] and might be exploited in the design of fluidic rectifiers and their applications.

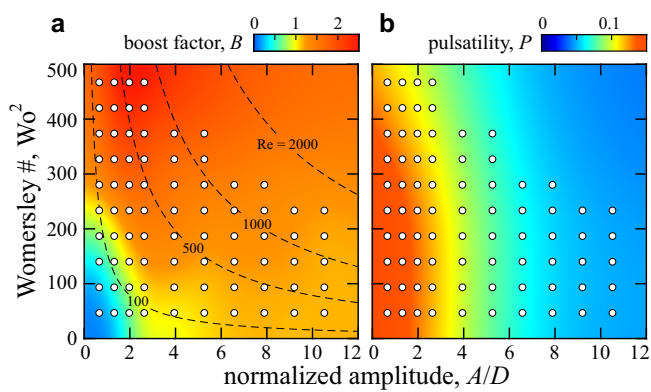

**Fig. 7 Comparison of pump rate and pulsatility in experiments and a quasi-steady model.** **a** Boost factor $B$ quantifying enhancement of pumping effectiveness in experiments relative to the model prediction. Markers indicate locations of measurements, and the colormap is interpolated and extrapolated elsewhere. **b** Pulsatility of the experimentally-measured rectified flow, defined as the ratio of ripple amplitude to mean flow rate in the DC branch.

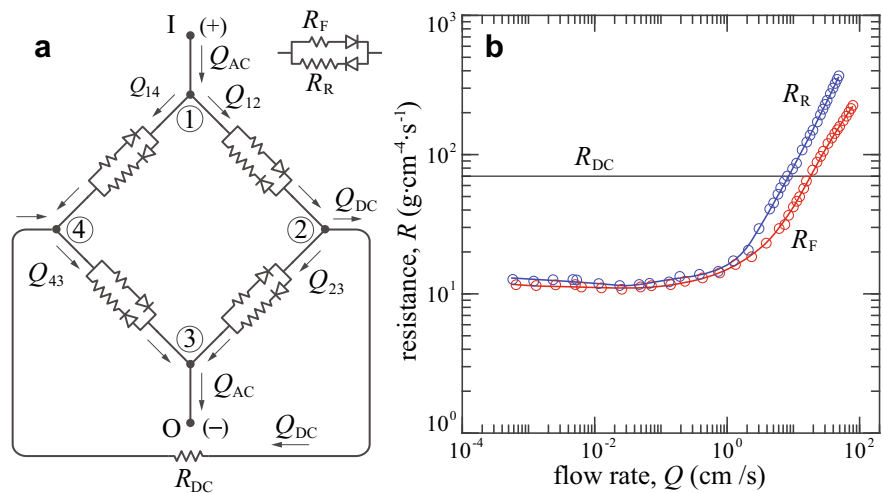

**Fig. 8 Circuit model with nonlinear resistances. a** Equivalent electronic circuit of the fluidic AC-to-DC converter, where the Tesla conduit is modeled as a 'leaky diode' with asymmetric resistance values. **b** Resistances vs. flow rate for forward (red) and reverse (blue) flow through Tesla's conduit, as inferred from the data shown in Fig. 2. The DC branch resistance value is estimated via the Hagen-Poiseuille law.

Towards interpreting the apparently early onset of turbulence, we first note that the system might be viewed as having qualities of both internal and external flows. This duality stems from the fact that the conduit width and its internal structures that 'invade' the flow are of the same scale. Said another way, the relative roughness is of order one. For external or open flow around a bluff body, intrinsic unsteadiness in the form of vortex shedding first appears at Re ≈ 100 and is accompanied by a transition from linear to quadratic scaling of drag with flow speed[6,42]. Analogous events would seem to occur for the local flows around the intruding structures in Tesla's device. Hence, the signatures of turbulence observed here may be viewed as arriving early from the standpoint of conventional internal or pipe flow but as expected for external flows.

The case of unsteady or pulsatile flow, while less studied, has been considered in previous experiments[11–13,15,33,63] and simulations[17,24] that demonstrate pumping using Tesla-inspired geometries. Our quantifications of performance offer concrete guidelines that might be generally useful for such applications: (1) Reynolds numbers should be kept in the hundreds or higher to fully activate diodic response; (2) The output pump rate can be expected to increase stronger than linearly in both the amplitude and frequency of driving pulsations; (3) Stronger AC driving and high frequencies in particular can be used to optimally exploit unsteady effects and boost pump rate; and (4) Stronger driving and high amplitudes in particular optimally exploit inertial effects for low-pulsatility output flow. The second and third properties seem to bear out Tesla's vision that diodicity is amplified for highly unsteady conditions[10]. Future work might assess these findings in simulations, test whether they are general features of asymmetric conduits[59–62], and optimize shape for pulsatile flow conditions.

While this work emphasizes fluid mechanical characterization rather than practicalities, many applications of Tesla-like valves have been proposed, investigated, and implemented[11–13,15–18,20,24,26–30,32–35,41,63]. Here we touch on two engineering contexts and their physiological analogs. First, the pumping applications discussed above speak to Tesla's original motivation of a no-moving-parts valve that is resistant to wear or failure. Our findings indicate that its valving action is best at high frequencies. This motivates the specific application in which the vibrations intrinsic to all forms of machinery are harnessed to pump coolants, fuels, lubricants, or other fluids needed for proper operation[59]. Here the leakiness or lossiness of the diode is no major detriment as the kinetic energy of the vibrational source is ample, ever-present and otherwise unused. This use is reminiscent of the return flow of blood up leg veins and of transport in the lymphatic system, both of which are driven by ambient muscle contractions that squeeze vessels and activate serial valvular structures[1,64,65]. Secondly, the device may be exploited for its direction-dependent flow states and especially the turbulent mixing available at unusually low Re. A fluid drawn in reversely can be readily mixed with other fluids, heated/cooled or otherwise conditioned or treated, after which it may be expelled for further processing or use. Biological analogs exist in respiratory systems, where inhaled air is heated and humidified by so-called turbinates in the nasal cavity[1,66] and where structures resembling Tesla valves have recently been discovered in the turtle lung[67].

In summary, our findings support two main conclusions: (1) For steady forcing, an abrupt turn-on of diodic behavior is linked to the transition to turbulent flow, which is triggered at anomalously low Re; and (2) For unsteady forcing, the diodic performance is enhanced several-fold and especially so for high-frequency pulsatile flow. In more detail, our steady pressure/flow-rate investigation reveals a transitional Reynolds number Re ≈ 200 marking the appearance of significant differences in the reverse versus forward resistances. This turn-on of diodicity is accompanied by the onset of nonlinear scaling of pressure drop with flow rate, departure from the laminar-flow friction law, and flow instabilities that induce unsteadiness and strong mixing. These steady-forcing results serve as a baseline for understanding unsteady performance, which we assess using an AC-to-DC converter that transforms forced oscillations into one-way flows. The system acts as a pump or rectifier whose output rate varies nonlinearly with input driving parameters. While significant pumping is observed only when the diodic response is activated, a quasi-steady model informed by steady-flow characteristics tends to underestimate the observed pump rates. The model also does not account for the smooth operation of the pump, as quantified by the low pulsatility of its output DC flow, nor the counter-intuitive decrease in DC pulsatility for stronger AC driving.

## Methods

**Circuit model for the AC-to-DC converter**. We model the AC-to-DC converter as the equivalent electronic circuit shown in Fig. 8a. Each Tesla conduit is represented by a combination of two ideal diodes and two resistances $R_F(Q)$ and $R_R(Q) \geq R_F(Q)$ whose values depend nonlinearly on flow rate. For terminals I and O connected to an oscillating source $Q_{AC}(t)$, we seek the time-averaged DC current $<Q_{DC}(t)>$ going from node 2 to node 4 through a resistor of constant resistance $R_{DC}$. We seek the solution by applying Kirchhoff's laws for nodes and loops: Total incoming flows and total outgoing flows are equal at a node, and the sum of all pressure drops around a closed loop is equal to zero. Ohm's law $\Delta p = QR$ describes the relationship of pressure $\Delta p$ across and flow rate $Q$ through any given resistor of resistance $R$.

In general, given a time-dependent input flow rate $Q_{AC}$, the resulting flow rates $Q_{12}$, $Q_{23}$, $Q_{43}$, $Q_{14}$, and $Q_{DC}$ also depend on time and thus so do $R_F$ and $R_R$. To simplify the notation, the time dependence of these quantities is suppressed in what follows. The number of unknown currents can be reduced by invoking symmetry. Consider when the source polarity is as indicated in Fig. 8a, that is, positive rate $Q_{AC}$ enters at I and exits at O. In the direction indicated by arrows, the sign of $Q_{DC}$ is positive, while the signs of $Q_{12}$, $Q_{43}$, $Q_{23}$, and $Q_{14}$ are not known a priori. Symmetry demands that $Q_{12} = Q_{43}$ and $Q_{23} = Q_{14}$. In the labelled directions, $Q_{12}$ and $Q_{14}$ must be like-signed (both positive or both negative) so that the net pressure drop around the closed loop 1-2-4-1 is zero. Further, in order for the flow rates to conserve at node 1, the sign in question must be positive. It then follows that all flow rates are positive in the labelled directions. There are then two independent equations corresponding to Kirchhoff's node law and one corresponding to the loop law:

$$\begin{aligned} Q_{12} - Q_{23} - Q_{DC} &= 0 \\ Q_{12} + Q_{23} - Q_{AC} &= 0 \\ Q_{12}R_F(Q_{12}) - Q_{23}R_R(Q_{23}) + Q_{DC}R_{DC} &= 0. \end{aligned} \quad (1)$$

Symmetry ensures that the same equations hold when the polarity of the source is reversed, i.e. $Q_{AC} \mapsto -Q_{AC}$. So the system of equations is valid in general and describes the circuit with an oscillating source at all instances in time. The algebraic system of 3 equations is nonlinear in the 3 unknowns $Q_{12}$, $Q_{23}$, and $Q_{DC}$, all of which depend on time, and involves the harmonic forcing $Q_{AC}(t) = Q_0 \sin(2\pi ft)$. Hence the system is periodic in time, so it is sufficient to solve and average over a single period $T = 1/f$. This together with the symmetry of the system under reversing $Q_{AC}$ as discussed above leads to the conclusion that it is only necessary to solve and average over a half-period. We solve the system numerically using MATLAB by stepping time through a half-period $t \in [0, T/2]$, where the exact functions for $R_F(Q)$ and $R_R(Q)$ are specified by spline fits to the experimental measurements, as shown in Fig. 8b.

Since the DC branch pipe has only a weak oscillatory component, we estimate its resistance $R_{DC}$ using Hagen-Poiseuille theory for developed and laminar pipe flow[6]. Accordingly, the resistance $R = 128\mu L/\pi D^4$ of a pipe is constant in time and depends on the fluid (water at room temperature) viscosity $\mu = 8.9 \times 10^{-3} \text{dyn} \cdot \text{s/cm}^2$ as well as the pipe length $L$ and diameter $D$. For experimental reasons, the DC branch consists of four different segments connected in series. Segments 1, 2, and 3 are pipes with different lengths (30, 150, 30 cm) and diameters (1, 2.4, 1 cm respectively), and segment 4 is a flow conditioner consisting of a bundle of 50 parallel straws, each 0.25 cm in diameter and 20 cm in length. Following the law of adding linear resistors, the resistance of segment 4 is 1/50 the resistance of each straw. We then estimate $R_{DC}$ as the sum of four resistances, $R_{DC} = (128\mu/\pi)(L_1/D_1^4 + L_2/D_2^4 + L_3/D_3^4 + L_4/20D_4^4) = 70 \text{ g} \cdot \text{cm}^{-4} \cdot \text{s}^{-1}$. At each time step, an instantaneous value of $Q_{DC}(t)$ is obtained using the MATLAB root-finding function fsolve. Averaging the solutions over the half-period $[0, T/2]$ yields the rectified flow rate $<Q_{DC}(t)>$. The effectiveness of rectification is then

$$E_M = \frac{\langle Q_{DC}(t) \rangle}{\langle |Q_{AC}(t)| \rangle} = \frac{\langle Q_{DC}(t) \rangle}{\frac{1}{T/2}\int_0^{T/2} Q_0 \sin(2\pi ft) \, dt} = \frac{\langle Q_{DC}(t) \rangle}{2Q_0/\pi} = \frac{U_{DC}}{4Af}, \quad (2)$$

which is consistent with the definition of $E$ given in the text in the context of the experiments. Here $Q_0 = (2\pi Af)wd$ and $\langle Q_{DC}(t) \rangle = \langle U_{DC}(t) \rangle wd = U_{DC}wd$, where $wd$ is equal to the average cross-sectional area of the conduit.

Though not true for Tesla's conduit, the simplified case of $R_F$ and $R_R$ being constants independent of flow rate gives some intuition for $E_M$. The system of equations (1) are then a linear system in the 3 unknowns $Q_{12}$, $Q_{23}$ and $Q_{DC}$. The output DC flow rate can be calculated analytically as

$$Q_{DC} = \frac{R_R - R_F}{R_R + R_F + 2R_{DC}} Q_{AC}, \quad (3)$$

which is valid at all instances for time-dependent flow rates $Q_{DC}(t)$ and $Q_{AC}(t)$. As in the full problem with rate-dependent resistance values, given the input $Q_{AC}(t) = Q_0 \sin(2\pi ft)$ and averaging over a half-period $t \in [0, T/2]$ yields

$$\langle Q_{DC}(t) \rangle = \frac{1}{T/2} \int_0^{T/2} \frac{R_R - R_F}{R_R + R_F + 2R_{DC}} Q_0 \sin(2\pi ft) \, dt = \frac{2Q_0}{\pi} \frac{R_R - R_F}{R_R + R_F + 2R_{DC}}. \quad (4)$$

Following Eq. (2), the effectiveness $E_M$ is then given by

$$E_M = \frac{R_R - R_F}{R_R + R_F + 2R_{DC}} = \frac{Di - 1}{Di + 1 + 2R_{DC}/R_F}, \quad (5)$$

where $Di = R_R/R_F$. Because $Di \geq 1$ and all resistance values are non-negative, $E_M \leq 1$. Some special cases and limits:

- For ideal diodes with $R_F = 0$ and $R_R = \infty$, $Di = \infty$ and $E_M = 1$, indicating perfect rectification.
- For symmetric resistors with $R_F = R_R$, $Di = 1$, and $E_M = 0$, indicating no rectification.
- If $R_{DC} \ll R_F \leq R_R$, then $E_M = (Di - 1)/(Di + 1)$ depends on the resistance values $R_R$ and $R_F$ only through their ratio Di.

## Data availability
All relevant data are available upon request to the authors.

## References

1. Nielsen, K. S. *Animal Physiology*, vol. 7 (Prentice-Hall, 1964).
2. Conway, A. *A Guide to Fluidics* (Macdonald/Elsevier, 1971).
3. Lal, J. *Hydraulic Machines including Fluidics* (Metropolitan Book Company, 1975).
4. Waite, L. & Fine, J. M. Applied biofluid mechanics (2007).
5. Fung, Y.-c. *Biomechanics: Mechanical properties of living tissues* (Springer, 2013).
6. Tritton, D. J. *Physical fluid dynamics* (Springer Science & Business Media, 2012).
7. Whitesides, G. M. The origins and the future of microfluidics. *Nature* **442**, 368–373 (2006).
8. Squires, T. M. & Quake, S. R. Microfluidics: Fluid physics at the nanoliter scale. *Rev. Mod. Phys.* **77**, 977 (2005).
9. Schlichting, H. & Gersten, K. *Boundary Layer Theory* (Springer, 2016).
10. Tesla, N. Valvular conduit (1920). US Patent 1,329,559.
11. Forster, F. K., Bardell, R. L., Afromowitz, M. A., Sharma, N. R. & Blanchard, A. Design, fabrication and testing of fixed-valve micro-pumps. *ASME-PUBLICATIONS-FED* **234**, 39–44 (1995).
12. Schlüter, M., Kampmeyer, U., Hermsdorf, A. & Lilienhof, H. A micro pump, compatible with multiple manufacturing methods, for fluidhandling in disposable microsystems. In *Micro Total Analysis Systems 2002*, 154–156 (Springer, 2002).
13. Gamboa, A. R., Morris, C. J. & Forster, F. K. Improvements in fixed-valve micropump performance through shape optimization of valves. *J. Fluids Eng.* **127**, 339 (2005).
14. Paul, F. W. Fluid mechanics of the momentum flueric diode. In *IFAC Symposium on Fluidics, Royal Aeronautical Society, Paper A*, vol. 1, 1–15 (1969).
15. Kord, J. et al. Hybrid synthetic jet actuator with a novel fluidic diode. *WIT Trans. Eng. Sci.* **74**, 493–503 (2012).
16. Zhang, S., Winoto, S. & Low, H. Performance simulations of tesla microfluidic valves. In *2007 First International Conference on Integration and Commercialization of Micro and Nanosystems*, 15–19 (American Society of Mechanical Engineers Digital Collection, 2007).
17. Anagnostopoulos, J. & Mathioulakis, D. Numerical simulation and hydrodynamic design optimization of a tesla-type valve for micropumps. *IASME Trans.* **2**, 1846–1852 (2005).
18. Forster, F. K. & Williams, B. E. Parametric design of fixed-geometry microvalves: The tesser valve. In *ASME 2002 International Mechanical Engineering Congress and Exposition*, 431–437 (American Society of Mechanical Engineers Digital Collection, 2002).

19. Bendib, S., Français, O., Tabeling, P. & Willaime, H. Analytical study and characterization of micro-channel and passive micro-diode. In *12th Micromechanics Europe Workshop*, 147–150 (2001).
20. Kim, J., Kang, K.-N., Jin, Y., Goettert, J. & Ajmera, P. K. Hydrodynamic focusing micropump module with pdms/nickel-particle composite diaphragms for microfluidic systems. *Microsyst. Technol.* 21, 65–73 (2015).
21. Nobakht, A., Shahsavan, M. & Paykani, A. Numerical study of diodicity mechanism in different tesla-type microvalves. *J. Appl. Res. Technol.* 11, 876–885 (2013).
22. Thompson, S. M., Paudel, B., Jamal, T. & Walters, D. Numerical investigation of multistaged tesla valves. *J. Fluids Eng.* 136, 081102 (2014).
23. Truong, T. & Nguyen, N. Simulation and optimization of tesla valves. *Nanotechnology* 1, 178–181 (2003).
24. Mohammadzadeh, K., Kolahdouz, E., Shirani, E. & Shafii, M. Numerical investigation on the effect of the size and number of stages on the tesla microvalve efficiency. *J. Mech.* 29, 527–534 (2013).
25. Ansari, S., Bayans, M., Rasimarzabadi, F. & Nobes, D. S. Flow visualization of the Newtonian and non-newtonian behavior of fluids in a tesla-diode valve. In *5th International Conference on Experimental Fluid Mechanics* (2018).
26. Arunachala, U., Rajat, A., Shah, D. & Sureka, U. Stability improvement in natural circulation loop using tesla valve–an experimental investigation. *Int. J. Mech. Prod. Eng. Res. Dev.* 9, 13–24 (2019).
27. Chandavar, R. A. Stability analysis of tesla valve based natural circulation loop for decay heat removal in nuclear power plants. In *2019 Advances in Science and Engineering Technology International Conferences (ASET)*, 1–6 (IEEE, 2019).
28. Hong, C.-C., Choi, J.-W. & Ahn, C. H. A novel in-plane passive microfluidic mixer with modified tesla structures. *Lab a Chip* 4, 109–113 (2004).
29. Hossain, S., Ansari, M. A., Husain, A. & Kim, K.-Y. Analysis and optimization of a micromixer with a modified tesla structure. *Chem. Eng. J.* 158, 305–314 (2010).
30. Yang, A.-S., Chuang, F.-C., Su, C.-L., Chen, C.-K. & Lee, M.-H. Development of a 3d-tesla micromixer for bio-applications. In *2013 IEEE International Conference on Mechatronics and Automation*, 152–157 (IEEE, 2013).
31. Thompson, S. M., Jamal, T., Paudel, B. J. & Walters, D. K. Transitional and turbulent flow modeling in a tesla valve. In *ASME 2013 International Mechanical Engineering Congress and Exposition* (American Society of Mechanical Engineers Digital Collection, 2013).
32. Wang, C.-T., Chen, Y.-M., Hong, P.-A. & Wang, Y.-T. Tesla valves in micromixers. *Int. J. Chem. React. Eng.* 12, 397–403 (2014).
33. Morris, C. J. & Forster, F. K. Low-order modeling of resonance for fixed-valve micropumps based on first principles. *J. Microelectromechanical Syst.* 12, 325–334 (2003).
34. De Vries, S., Florea, D., Homburg, F. & Frijns, A. Design and operation of a tesla-type valve for pulsating heat pipes. *Int. J. Heat. Mass Transf.* 105, 1–11 (2017).
35. Forster, F. K. & Walter, T. Design optimization of fixed-valve micropumps for miniature cooling systems. In *ASME 2007 InterPACK Conference collocated with the ASME/JSME 2007 Thermal Engineering Heat Transfer Summer Conference*, 137–145 (American Society of Mechanical Engineers Digital Collection, 2007).
36. Deng, Y., Liu, Z., Zhang, P., Wu, Y. & Korvink, J. G. Optimization of no-moving part fluidic resistance microvalves with low Reynolds number. In *2010 IEEE 23rd International Conference on micro electro mechanical systems (MEMS)*, 67–70 (IEEE, 2010).
37. Lin, S., Zhao, L., Guest, J. K., Weihs, T. P. & Liu, Z. Topology optimization of fixed-geometry fluid diodes. *J. Mech. Design* 137 (2015).
38. Abdelwahed, M., Chorfi, N. & Malek, R. Reconstruction of tesla micro-valve using topological sensitivity analysis. *Adv. Nonlinear Anal.* 9, 567–590 (2019).
39. Stone, H. Fundamentals of fluid dynamics with an introduction to the importance of interfaces. *Soft Interfaces: Lect. Notes Les. Houches Summer Sch.: Vol. 98, July 2012* 98, 1 (2017).
40. Magoon, D., Sternthal, J., Tremml, W. & Epps, B. Poster: Visualization of tesla's valvular conduit (2017).
41. Porwal, P. R., Thompson, S. M., Walters, D. K. & Jamal, T. Heat transfer and fluid flow characteristics in multistaged tesla valves. *Numer. Heat. Transf., Part A: Appl.* 73, 347–365 (2018).
42. White, F. M.*Fluid mechanics* (WCB/McGraw-Hill, 1999).
43. Martin, H. The generalized lévêque equation and its practical use for the prediction of heat and mass transfer rates from pressure drop. *Chem. Eng. Sci.* 57, 3217–3223 (2002).
44. Moody, L. F. Friction factors for pipe flow. *Trans. Asme* 66, 671–684 (1944).
45. Bellos, V., Nalbantis, I. & Tsakiris, G. Friction modeling of flood flow simulations. *J. Hydraulic Eng.* 144, 04018073 (2018).
46. Kandlikar, S. G., Schmitt, D., Carrano, A. L. & Taylor, J. B. Characterization of surface roughness effects on pressure drop in single-phase flow in minichannels. *Phys. Fluids* 17, 100606 (2005).
47. Nikuradse, J. *Laws of flow in rough pipes*, vol. 2 (National Advisory Committee for Aeronautics Washington, 1950).
48. Li, L., Mo, J. & Li, Z. Nanofluidic diode for simple fluids without moving parts. *Phys. Rev. Lett.* 115, 134503 (2015).
49. Pope, S. B. Turbulent flows (2001).
50. Horowitz, P. & Hill, W.*The Art of Electronics* (Cambridge University Press, 1989).
51. Eguchi, T., Watanabe, S., Takahara, H. & Furukawa, A. Development of pulsatile flow experiment system and piv measurement in an elastic tube. *Mem. Fac. Eng., Kyushu Univ.* 63, 161–172 (2003).
52. Becker, A. D., Masoud, H., Newbolt, J. W., Shelley, M. & Ristroph, L. Hydrodynamic schooling of flapping swimmers. *Nat. Commun.* 6, 1–8 (2015).
53. Ramananarivo, S., Fang, F., Oza, A., Zhang, J. & Ristroph, L. Flow interactions lead to orderly formations of flapping wings in forward flight. *Phys. Rev. Fluids* 1, 071201 (2016).
54. Newbolt, J. W., Zhang, J. & Ristroph, L. Flow interactions between uncoordinated flapping swimmers give rise to group cohesion. *Proc. Natl Acad. Sci. USA* 116, 2419–2424 (2019).
55. Thielicke, W. & Stamhuis, E. Pivlab–towards user-friendly, affordable and accurate digital particle image velocimetry in matlab. *J. Open Res. Softw.* 2 (2014).
56. Womersley, J. R. Method for the calculation of velocity, rate of flow and viscous drag in arteries when the pressure gradient is known. *J. Physiol.* 127, 553–563 (1955).
57. Loh, S. & Blackburn, H. Stability of steady flow through an axially corrugated pipe. *Phys. Fluids* 23, 111703 (2011).
58. Cotrell, D., McFadden, G. B. & Alder, B. Instability in pipe flow. *Proc. Natl Acad. Sci. USA* 105, 428–430 (2008).
59. Thiria, B. & Zhang, J. Ratcheting fluid with geometric anisotropy. *Appl. Phys. Lett.* 106, 054106 (2015).
60. Groisman, A. & Quake, S. R. A microfluidic rectifier: anisotropic flow resistance at low reynolds numbers. *Phys. Rev. Lett.* 92, 094501 (2004).
61. Hawa, T. & Rusak, Z. Viscous flow in a slightly asymmetric channel with a sudden expansion. *Phys. Fluids* 12, 2257–2267 (2000).
62. Fadl, A., Zhang, Z., Faghri, M., Meyer, D. & Simmon, E. Experimental investigation on geometric effect on micro fluidic diodicity. In *International Conference on Nanochannels, Microchannels, and Minichannels*, vol. 4272, 489–492 (2007).
63. Thompson, S., Ma, H. & Wilson, C. Investigation of a flat-plate oscillating heat pipe with tesla-type check valves. *Exp. Therm. Fluid Sci.* 35, 1265–1273 (2011).
64. Buxton, G. A. & Clarke, N. Computational phlebology: the simulation of a vein valve. *J. Biol. Phys.* 32, 507–521 (2006).
65. Schmid-Schonbein, G. W. Microlymphatics and lymph flow. *Physiological Rev.* 70, 987–1028 (1990).
66. Keyhani, K., Scherer, P. & Mozell, M. Numerical simulation of airflow in the human nasal cavity. *J. Biomech. Eng.* 117, 429–441(1995).
67. Farmer, C., Cieri, R. & Pei, S. A tesla valve in a turtle lung: Using virtual reality to understand and to communicate complex structure-function relationships. In *Journal of Morphology*, vol. 280, S70–S70 (Wiley, 2019).

### Acknowledgements
We thank S. Childress, M. Shelley, and J. Zhang for useful discussions. We acknowledge support from the National Science Foundation (DMS-1646339 and DMS-1847955 to L.R.).

### Author contributions
Q.M.N. and L.R. designed the experiments, formulated the model, and wrote the paper. Q.M.N., J.A., and L.R. contributed to conducting the experiments, analyzing the data, interpreting the results, and reviewing the manuscript.

### Competing interests
The authors declare no competing interests.

### Additional information
**Supplementary information** The online version contains supplementary material available at https://doi.org/10.1038/s41467-021-23009-y.

**Correspondence** and requests for materials should be addressed to L.R.

**Peer review information** *Nature Communications* thanks Bin Liu and Scott Thompson for their contribution to the peer review of this work. Peer reviewer reports are available.

**Reprints and permission information** is available at http://www.nature.com/reprints

**Publisher's note** Springer Nature remains neutral with regard to jurisdictional claims in published maps and institutional affiliations.